# Addressing Packet Dropping Misbehavior using Metaheuristic Approach: A Survey

*Kishan N.Patel*
*PG Scholar,Department of Information Technology*
SVM Institute of Technology
Bharuch 392-001, Gujarat, India
knpatel5900@yahoo.com

*Rutvij H.Jhaveri*
*Asst.professor,Department of Information Technology*
SVM Institute of Technology
Bharuch 392-001, Gujarat, India
rhj_svmit@yahoo.com

***Abstract*— A Mobile Ad hoc Network can be said as a collection of mobile nodes, which builds a dynamic topology and a resource constrained network. Due to dynamic nature routing is challenging in MANET. In Mobile Ad hoc Networks different type of attack is possible. By manipulation in routing and attacks against data are reasoned by malicious nodes. MANETs are susceptible to various attacks from selfish nodes. To defeat those problems security is demanded for preventing node from various attacks. Metaheuristic techniques are used for addressing a problem found in ad hoc networks such as routing, security and optimization. A lot of research has been carried out for secure routing process with metaheuristic approaches. In this paper, we present survey of metaheuristic approach to improve routing process by enhancing the security and optimization among different nodes in mobile ad hoc networks.**

## I. Introduction

MOBILE AD HOC NETWORK (MANET) [1] is a self configuring infrastructure less network in which the mobile devices are connected by wireless medium. In MANET devices are free to move independently in any direction, and can change its links to any other devices in the network frequently. The phenomenon of movement of the data packet from source to destination is known as routing which means that the process of selecting best paths in a network. A routing protocol uses routing algorithms as well as software to find an optimal path in network for transfer and communication between network nodes.

A number of systems in nature and society are modeled by networks with complex topology. One important aspect of a network is its capability to withstand failures and fluctuations in the operation of its nodes and links. Failure may occur in different ways, depending on the complexity of the system. There are many ways to define the robustness of a network, a network is robust if their function is not affected by the attacks to nodes or links, which can be either random or malicious. Thus, the robustness of networks is to guarantee the security of network systems, World Wide Web etc. Therefore, the security for network structures using metaheuristic approach has been studied in this paper.

This paper is organized as follows. Section-II highlights the Metaheuristic . The Metaheuristic algorithms in section-III. Section-IV discusses related work. The conclusion is presented in section-VI.

## II. Metaheuristic

Heuristic [2] means to find or to discover by trial and error. And meta means higher level and metaheuristics generally perform better than simple heuristics. The word "metaheuristic” [3] can be considered as a "master strategy that guides and modifies other heuristics to produce solutions. Generally metaheuristic is used for solving problem in ad hoc networks. There are two techniques: Online metaheuristic approach and Offline metaheuristic approach. The main difference between them is the moment when they are applied for solving problem. Online metaheuristic approaches are used for decision making or solving the problem during run time. Offline metaheuristic approaches are useful when there is no requirement for system to adapt change during runtime.

### *A. Properties of Metaheuristic [4]*

1) Search process is influenced by Metaheuristic approch.
2) The main purpose of the search process is to find near to optimal solutions.
3) It range from simple local search procedures to complex processes.
4) They are non-deterministic.
5) They are not problem-specific.

### *B. Why Metaheuristic?*

1) Metaheuristics are used for solving the security and routing problems in ad hoc networks.
2) Security problem in the network may be due to selfish and malicious node. Metaheuristic approach can be used for overcoming the problem of node misbehaving in ad hoc networks.
3) For route optimization e.g selecting the shortest and quality route from source to destination in the ad hoc networks.
4) Metaheuristic approach is also used for solving the supply chain problems.
5) Metaheuristic approach is important for generating optimum set of test data in software testing.

## III. METAHEURISTIC ALGORITHMS

Metaheuristic [5] is a procedure for finding or select a lower heuristic and provide a good solution to an optimization and security problem with limited computation capacity. Metaheuristics may make few assumptions about the problems in ad hoc networks so it is used for a variety of problems.

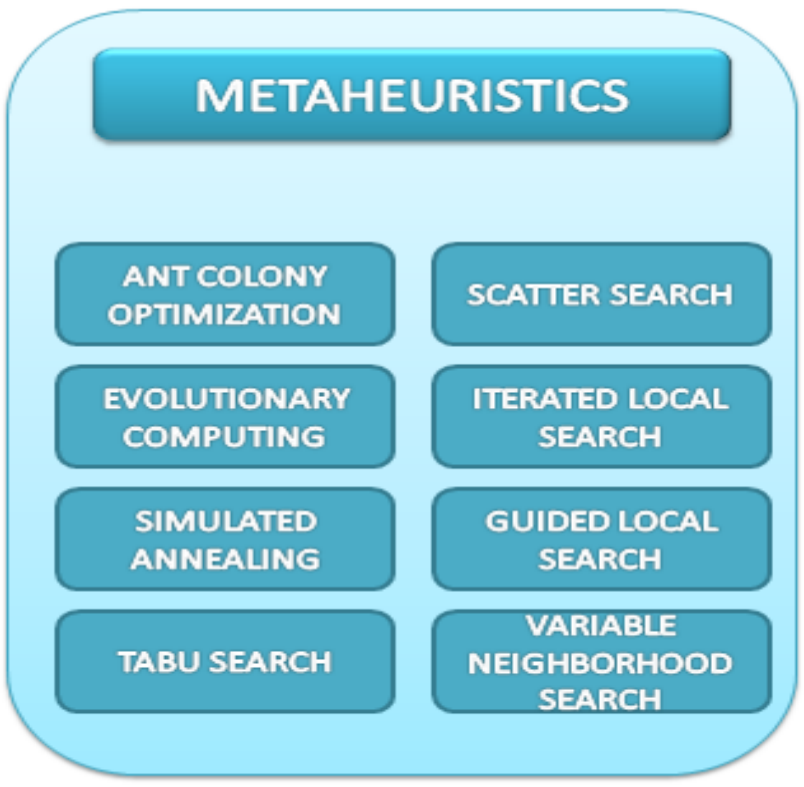

Fig.1. Classification of Metaheuristics [6]

As shown in above fig.no.1 classification of metaheuristics. It includes Ant colony optimization (ACO), Tabu search (TS), scatter search(SS), Variable neighborhood search (VNS), Guided local search (GLS), iterated local search (ILS), simulated annealing (SA), evolutionary algorithms (EC) etc.

Ant colony optimization (ACO) [7] is a population-based metaheuristic for the solution to the difficult combinatorial optimization problems. The inspiring source of ACO is behavior of real ants. This behavior empowers ants to find shortest paths from food sources to their nest. While walking from food sources to the nest, ants dregs a substance called pheromone on the ground. This behavior is the basis for a cooperative interaction for emerging of shortest paths.

Tabu search [8] is a heuristic procedure for solving optimization problems. Tabu search has obtained optimal and near to optimal solutions to a wide variety of practical problems in varied applications such as scheduling, character recognition.

Scatter Search SS [9] is designed to operate on a set of points, which is called reference points. It also captures information which is not present in reference points. It takes advantage of auxiliary heuristic methods.

Variable neighbourhood search VNS [10] is based on simple principal . It starts from feasible solution X and from a predefined set of steps, chooses a random solution within the neighborhood of X and move there if the solution is better.

Guided local search GLS [11] is penalty based metaheuristic which is on the top of another search algorithm, which aims to improve their efficiency and robustness.

In Iterated local search ILS [12] one iteratively builds a sequence of solutions generated by the embedded heuristic, leading for better solutions than use repeated random trials of that heuristic.

Simulated annealing [13] is a probabilistic metaheuristic for the global optimization problem for finding a good approximation to the global optimum for a given function in a large search space.

Evolutionary computing [14], Evolution is a process in which functioning deals with chromosomes. The process takes place during the reproduction stage. For e.g mutation and recombination. It is combination of genetic algorithms, evolution strategies, evolutionary programming and genetic programming**.**

## IV. RELATED WORK

Most of the literature done on metaheuristics is experimental in nature. This section describes the several Metaheuristic approaches used for gathering the knowledge for establishing the security and optimization.

Singh et al. [15] proposed a new ACO based Routing Algorithm(ANTALG). There will be random selection of source and destination nodes and exchanges the Ants between them. And also optimal behavior is not for searching shortest-hop paths but also for the quality of the links which make up those paths. Future packets uses those in a stochastic manner.

Srinivas et al. [16] proposed FTAR utilized Ant-Colony Optimization (ACO) which is the food-searching algorithm of real ant-agent. Fuzzification utilized the time-ratio which is ratio between the route-reply time and time-to-live. It withal access dropped packet parameter utilized to measure the number of packet dropped at node. FTAR uses two types of control packet: BANT and FANT.

Bahaa et al. [17] proposed an Intelligent Ant Colony Multi Agent based Trusted on demand Routing Protocol for MANETs called TARA. Trust value for each node based on self-monitoring. This protocol is to minimize the number of messages to be exchanged. The trust value propagation is eliminated in this protocol. Ant Colony Optimization is used to find the optimal route resulting in a better performance and lower latency. In the presence of malicious nodes TARA works better other trusted routing protocol.

Ahmed et al. [18] LBMRAA delivers more packets to the destination in less time and with lower overhead. It uses to found all the routes for data delivery which distributes the load over any possible route to destination where the more reliable and suitable routes get more traffic than less reliable and suitable routes.

Alrajeh et al. [19] It is based on ant colonization technique. It uses hello packets for discovering surrounding neighbors. It uses two paths for to overcome the problem of node failure but also to increase the efficiency of overall network. Forward ants are for increment the reputation values along the path. Destination node uses backward ants which carry information and instruction about route security, it shows better performance in, routing overheads, and data forwarding efficiency. In the presence of malicious node it shows high data delivery rate.

Melchor et al. [20] proposed AntTrust it facilitates a malicious manipulations of data packets. The protocol is for ensuring routing features and for adding some security level to routes and inspired by multi-agent systems by ant colony and their ability to solve complex problems for establishing paths in ad-hoc networks.

Satpute et al. [21] Particle swarm optimization combined with Machine Learning techniques for Anomaly Detection in Network Intrusion Detection System and to enhance the performance of intrusion detection system.

Sreelaja et al. [22] Proposed an Ant Colony Optimization Attack Detection (ACO-AD) algorithm for detecting sink hole attack and Ant Colony Optimization Boolean Expression Evolver Sign Generation (ABXES) for distributing key among the group of node and identifies anomalous connection.

Ranjan Sahoo et al. [23] proposed a new clustering model based on the concept of Trust mechanism. Ant colony routing establishes efficient communication between VANET nodes. Trust in the routing algorithms in terms of routing overhead. 20 random values are obtained for different number of VANET nodes and clusters and average value is presented.

Nancharaiah et al. [24] proposed algorithms is hybridization of PSO with ACO for MANETs PSO_ACO algorithm has an ability to cope with huge networks with large number of nodes. This hybrid algorithm has less total link delay and least communication cost compared with conventional ACO. This hybrid algorithm is very useful in mobile communications.

Simaremare et al.[25] proposed a new trust mechanism for security against DOS attacks. And also used ant colony optimization in which ants can move freely to find their destination. It will deposit pheromones if node is trusted. By using ACO throughput and PDR is increased.

Zafar et al. [26] First phase of the protocol uses genetic algorithm to provide QOS by selecting the fittest shortest route among the number of routes to provide optimization. Confidentiality and Authentication of packets for routing is achieved in second phase by trust based. Trust based packet delivering scheme is used for detecting and isolating the malignant nodes by using the routing layer information. The decision of the neglecting the node is done on the basis of the trust weigh. If the trust weigh value falls below a trust threshold, the equivalent node is marked as malignant.

Kaur et al. [27] Integrates ACO and DRI tables and verifies with AODV routing protocol. Promiscuous mode is use to check DRI tables of node. It generates and receive all traffic through node. DRI table contains dynamic routing information Node can be said as normal when it passes all traffic. Value from table is 1 when node is normal and if node is malicious then the value from table will 0. We will use DRI tables to check whether the node is malicious or not.

Chun Lo et al. [28] proposed ATBRA algorithm it is an ant based algorithm combined with Tabu list. It mitigates the selfish nodes problems occurring during route discovery phase and these kinds of algorithms are more suitable for dynamic networks algorithm. Under the dynamic topology and selfish nodes conditions results of SDR and RO simulation is efficient.

Jing Liu et al. [29] memetic algorithm, is proposed to enhance the robustness of an existing scale-free network against malicious attacks. Global search operator such as crossover operator, and local search operator, are designed to solve the problem of optimizing network structure. Also shown that real-world networks show the good performance of MA–RSFMA over existing local search methods.

Sherin Zafar et al. [30] proposed biometric stationed authentication protocol (BSAP) with meta-heuristic genetic algorithm**.** First phase create strong biometric features said as fingerprint that generates fittest offspring. Genetic operations are done to overcome the shortcomings of biometrics. A biometric cryptographic key is obtain for security in ad hoc networks which develops trust between various nodes of ad-hoc network.

Holloway et al. [31] evaluated (self organized multi agent swarm design) SOMAS performance and analysis of Algorithm Effectiveness, behavior Effectiveness, behavior Self Organization and Pareto Front Analysis. Testing is required for different agent rules with different scenario. The self organization metric should be used to determine the influence of self organization on the swarm’s effectiveness and efficiency.

Kalucha et al. [32] It contains four main phases: 1. Initialization Phase: The food sources are randomly generated by scout bees. 2. Employed Bee Phase: Employed bee finds a new food source within the neighborhood of the food source. 3. Selection Phase of Onlooker Bees. In these Onlooker bees calculates the profitability of food sources. Each onlooker bee selects one of the food sources based on the fitness value 4. Scout Bee Phase. It checks if there is any exhausted source to be abandoned. Food source is abandoned if position cannot be improved in number of cycles.

## Conclusion

The metaheuristic approach is currently attracting considerable interest for research community to meet the rising need. The important advantage of metaheurisic approach is to provide provably optimal solutions but they have potential to produce good solutions in short amounts of time. In this paper we present a survey of various metaheuristic algorithms and its functioning to overcome security and optimization problem to improve the performance of ad hoc networks. In order to overcome the shortcoming of existing algorithms and to enhance the robustness of an existing network against malicious attacks various techniques are hybridized. We conclude that metaheuristic approach can provide a best solution for security and optimization problem with limited computation.

Appendix A describes the survey on various Metaheuristic approaches with novel conceptions.

TABLE I: A survey on various Metaheuristic approaches

| Authors, Reference | Title | Mechanism / Algorithm | Purpose | Methodology |
|---|---|---|---|---|
| Singh et al.[15] | ANTALG: An Innovative ACO based Routing Algorithm for MANETs. | ACO based Routing Algorithm (ANTALG) | For reducing Packet dropping and average End-to-End delay. | There will be a random selection of source and destination nodes and exchanges the Ants(agents) between them. Optimal behavior is not merely searching shortest-hop path, but it also considers the quality of the links which make up those paths. |
| srinivas et al.[16] | Fuzzy-based trusted ant routing (FTAR) protocol in mobile ad hoc networks. | Fuzzy-based trusted ant routing (FTAR) | Optimal path, trusted routing. | Distinguish the healthy and malicious nodes by utilizing the parameters: time-ration and dropped packet. |
| Bahaa et al.[17] | TARA: Trusted Ant Colony Multi Agent Based Routing Algorithm for Mobile Ad-Hoc Networks. | TARA: Trusted Ant Colony Multi Agent Based Routing Algorithm | To avoid trust value propagation. To minimize the number of messages been exchanged. To find the best route for delivery. | Trust value of each node is directly applied to the route and no need to propagate the trust values like other trusted protocols. |
| Ahmed et al.[18] | Ant Colony and Load Balancing Optimizations for AODV Routing Protocol | Multi-Route AODV Ant routing (MRAA) Load balancing( LBMRAA) | To reduce the routing overhead, buffer overflow, end-to-end delay and increase the performance | Data packets are balanced over discovered paths and energy consumption is distributed across many nodes through network. |
| Alrajeh et al. [19] | Secure Ant-Based Routing Protocol for Wireless Sensor Network. | Secure Ant-Based Routing Protocol(SARP) | To provide Route security in network. | It uses two paths for data forwarding to overcome the problem of node failure and to increase the efficiency of overall network. |
| Melchor et al.[20] | AntTrust: A Novel Ant Routing Protocol forWireless Ad-hoc Network Based on Trust Between Nodes. | Ant Trust Routing Protocol | To increase the security of route. And malicious manipulations of data packets. | AntTrust is located precisely in the context of the security of routing. It also facilitates malicious manipulations of data packets. |
| Satpute et al.[21] | A Survey on Anomaly Detection in Network Intrusion Detection System Using Particle Swarm Optimization Based Machine Learning Techniques. | Particle swarm optimization combined with Machine Learning techniques for Anomaly Detection in Network Intrusion Detection System. | To enhance the performance of intrusion detection system. | Recently developed IDS with single techniques is insufficient against increasing threats in the network. So hybridization of techniques are used to satisfy the increasing demand of intelligent Intrusion Detection System (IDS). |
| Sreelaja et al.[22] | Swarm intelligence based approach for sinkhole attack detection in wireless sensor networks. | Proposed an Ant Colony Optimization Attack Detection (ACO-AD) algorithm and Ant Colony Optimization Boolean Expression Evolver Sign Generation (ABXES). | To detect a sinkhole attack and to identify an intruder in a wireless sensor networks. | ACO-AD algorithm for detection of sink hole attack, and ABXES algorithm for distributing keys among the group of node. And for identification of the anomalous connections without generation of false positives and minimization of storage in the sensor nodes. |
| Ranjan Sahoo et al.[23] | A trust based clustering with ant colony routing in Vanet. | Proposed a new Mobility-aware Ant Colony Optimization Routing (MAR-DYMO). | To decrease the routing overhead by establishing creating trust in between the nodes in vanet. | As number of vehicle increases when routing overhead increases. Trust in the routing algorithms in terms of routing overhead. 20 random values are obtained for different number of VANET nodes and clusters and the average value is presented. |
| Nancharaiah et al.[24] | MANET link Performance using Ant Colony Optimization and Particle Swarm Optimization Algorithms. | Ant colony optimization and Particle swarm optimization. | Finds the best solution over the particle's position and velocity with the cost and minimum. End-to-end delay. | Ant Colony Optimization (ACO) algorithm uses mobile agents as ants to discover feasible and best path in a network and PSO finds the best solution over the particle's position and velocity with the objective of cost and minimum End-to-end delay. |
| Simaremare et al.[25] | Performance analysis of optimized Trust AODV using ant Algorithms | Ant colony optimization | To sustain security against the dos attacks. | At agents can move freely to find destination it will update positive pheromone to the routing table. The pheromone is deposited if node is trusted. |
| Zafar et al.[26] | Trust Based QOS Protocol(TBQP) using Metaheuristic Genetic Algorithm for Optimizing and Securing MANET | Trust Based QOS Protocol (TBQP) Using Meta-heuristic Genetic Algorithm.. | To provide QOS by selecting the fittest shortest route among the number of routes to provide optimization. Acquaintance And Authentication of packets for routing in network. | Intriguing a trust based packet delivering scheme for detecting and isolating the malignant nodes using the routing layer information. A trust weigh is maintained and a node is remunerated by decreasing or increasing the trust weigh value. If the trust weigh falls below a trust threshold, node is marked as malicious node. |
| Kaur et al.[27] | A novel approach to prevent blackhole attack in wireless sensor network | Ant colony optimization | To detect and prevent black hole attack in MANET. | Integration of ACO and DRI tables and verification with AODV routing protocol. Node can be said as normal when it passes all traffic. Value from table is 1 when node is normal and if node is malicious then the value from table will 0. DRI tables to check whether the node is malicious or not. |
| Chun Lo et al.[28] | Mitigating Routing Misbehavior Using Ant-Tabu-Based Routing Algorithm for Wireless Ad-Hoc Networks | Ant-Tabu-Based Routing Algorithm(ATBR) | To increase successful delivery rate (SDR). To decrease routing overhead (RO). To prevent from dos attacks to enhance network performance. | It integrates both the advantages of both DSR and Ant-Colony Based Routing Algorithm (ARA). first two phases finds a routing path and TS with short-term memory is used to avoid getting stuck into local optimal and exclude the selfish nodes. |
| Jing Liu et al.[29] | A memetic algorithm for enhancing the robustness of scale-free networks against malicious attacks. | Memetic algorithm | To enhance the robustness of networks against malicious attacks. | By using Hill climbing, Simulated annealing, Smart rewiring and memetic algorithms for solving the problem of optimizing and security in network . |
| Zafar et al.[30] | Sustaining Security in MANET: Biometric Stationed Authentication Protocol (BSAP) Inculcating Meta-Heuristic Genetic Algorithm. | Biometric Stationed Authentication Protocol (BSAP) | To sustain security in MANET from DOS attacks. | Biometric cryptographic key is produced to enhance security of MANET. Hence data is protected by applying three levels of security by our prospective approach which develops trust between various nodes of ad-hoc network. |
| Holloway et al.[31] | Network Security Using Self Organized Multi Agent Swarms. | Self organized multi agent swarms (SOMAS) | To provides a novel computer network security management framework. | It represents the individual agents and their interaction with each other as well as the generic computer network environment. Each agent has a set of decision rules which is used to determine whether to execute a given actuator based on its observation of its state and environment and basic weighted discriminant function is used. |
| Kalucha et al.[32] | A Review on Artificial Bee Colony in MANET | Artificial Bee Colony (ABC) | To solve optimization problems, maximizes the lifetime of network. And also provides effective multi path data transmission. | ABC algorithm uses simple control parameters, it solves multimodal and multidimensional optimization problems. It can also applied to combinatorial and functional optimization problems |